\documentclass[fleqn,10pt]{wlscirep}
\usepackage[utf8]{inputenc}
\usepackage[T1]{fontenc}
\usepackage{float}
\usepackage[warnundef]{jabbrv}

\title{The formation of $30\,M_\odot$ merging black holes at solar metallicity}
\author[1,2,*]{Simone\,S.\,Bavera}
\author[1,2]{Tassos\,Fragos}
\author[3]{Emmanouil\,Zapartas}
\author[4,5]{Jeff\,J.\,Andrews}
\author[5]{Vicky\,Kalogera}
\author[6]{Christopher\,P.\,L.\,Berry}
\author[1,2]{Matthias\,Kruckow}
\author[5]{Aaron\,Dotter}
\author[1,7,8]{Konstantinos\,Kovlakas}
\author[1]{Devina\,Misra}
\author[5]{Kyle\,A.\,Rocha}
\author[5,9]{Philipp\,M.\,Srivastava}
\author[5]{Meng\,Sun}
\author[1,2]{Zepei\,Xing}

\affil[1]{Département d’Astronomie, Université de Genève, Chemin Pegasi 51, CH-1290 Versoix, Switzerland}
\affil[2]{Gravitational Wave Science Center (GWSC), Université de Genève, CH-1211 Geneva, Switzerland}
\affil[3]{IAASARS, National Observatory of Athens, Vas. Pavlou and I. Metaxa, Penteli, 15236, Greece}
\affil[4]{Department of Physics, University of Florida, 2001 Museum Rd, Gainesville, FL 32611, USA}
\affil[5]{Center for Interdisciplinary Exploration and Research in Astrophysics (CIERA) and Department of Physics and Astronomy, Northwestern University, 1800 Sherman Ave, Evanston, IL 60201, USA} 
\affil[6]{SUPA, School of Physics and Astronomy, University of Glasgow, Glasgow, G12 8QQ, UK}
\affil[7]{Institute of Space Sciences (ICE, CSIC), Campus UAB, Carrer de Magrans, 08193 Barcelona, Spain}
\affil[8]{Institut d’Estudis Espacials de Catalunya (IEEC), Carrer Gran Capit\`a, 08034 Barcelona, Spain}
\affil[9]{Electrical and Computer Engineering, Northwestern University, 2145 Sheridan Road, Evanston, IL 60208, USA}

\affil[*]{e-mail: Simone.Bavera@unige.ch}

\begin{abstract}
\textbf{
The maximum mass of black holes formed in isolated binaries is determined by stellar winds and the interactions between the binary components. 
We consider for the first time fully self-consistent detailed stellar structure and binary evolution calculations in population-synthesis models and a new, qualitatively different picture emerges for the formation of black-hole binaries, compared to studies employing rapid population synthesis models.
We find merging binary black holes can form with a non-negligible rate ($\sim 4\times10^{-7}\,M_\odot^{-1}$) at solar metallicity.
Their progenitor stars with initial masses $\gtrsim 50\,M_\odot$ do not expand to supergiant radii, mostly avoiding significant dust-driven or luminous blue variable winds. Overall, the progenitor stars lose less mass in stellar winds, resulting in black holes as massive as $\sim 30\,M_\odot$, and,
approximately half of them avoid a mass-transfer episode before forming the first-born black hole. 
Finally, binaries with initial periods of a few days, some of which may undergo episodes of Roche-lobe overflow mass transfer, result in mildly spinning first-born black holes,  $\chi_\mathrm{BH1} \lesssim 0.2$, assuming efficient angular-momentum transport. 
}
\end{abstract}
\begin{document}

\flushbottom
\maketitle
%\thispagestyle{empty}
%\section*{Introduction}
\vspace{-0.5cm}
Stellar-mass black holes (BHs) are known to populate our Universe. Evidence of their existence comes from X-ray binary (XRB) observations,\cite{2006ARA&A..44...49R,2016A&A...587A..61C,2016ApJS..222...15T} gravitational-wave (GW) observations\cite{2016PhRvL.116f1102A,2019PhRvX...9c1040A,2021PhRvX..11b1053A,2021arXiv210801045T,2021arXiv211103606T} of coalescing binary BHs (BBHs), and, more recently, from joint radial velocity and astrometric observations.\cite{2023MNRAS.518.1057E} 
The inferred BH masses from XRBs and astrometric binaries cover the range from $4.5\,M_\odot$ to $21\,M_\odot$,\cite{2016A&A...587A..61C,
%2010ApJ...725.1918O,2011ApJ...741..103F, %not observational papers
2021Sci...371.1046M,2023MNRAS.518.1057E} while BHs observed through GWs the range from $2.6\,M_\odot$ to $106\,M_\odot$\cite{
2021arXiv210801045T,2021arXiv211103606T
%2020ApJ...896L..44A,2020PhRvL.125j1102A
} (assuming the lower-mass object in GW190814 is a BH). 
The dynamical measurement of BH masses in XRBs and astrometric binaries is possible only in the Milky Way and in nearby galaxies for XRBs,\cite{2016ApJS..222...15T} while current detections of GWs probe the Universe up to redshift $z\sim 1$.\cite{2021arXiv210801045T,2021arXiv211103606T}
The discrepancy between the maximum BH mass inferred from XRBs and GWs has led to the belief that BHs with masses of around $30\,M_\odot$
%(below the pair-pulsation-instability regime) 
can only originate from the evolution of massive stars born with sub-solar metallicities,\cite{2010ApJ...714.1217B,2015MNRAS.451.4086S,2016PhRvL.116f1102A,2016Natur.534..512B,2016ApJ...818L..22A} namely below $Z_\odot = 0.0142$.\cite{2009ARA&A..47..481A} 
This belief is anchored in the empirical evidence that stellar-wind mass-loss rates for massive stars scale with increasing metallicity.\cite{2001A&A...369..574V,2021MNRAS.504.2051V,1988A&AS...72..259D,2000A&A...360..227N}
%and the fact that the most massive BH ever observed in the Milky Way from a HMXB, Cygnus X-1\cite{2021Sci...371.1046M}, is well below this limit. 
The maximum BH mass predicted by single stellar models depends on the mass loss of massive stars; however, this is uncertain due to the lack of observations of massive stars during their late evolutionary phases. \cite{1994A&AS..103...97M,2014ARA&A..52..487S,2017A&A...603A.118R,2020ApJ...890..113B,2021arXiv210908164V}
%However, the assumed stellar wind mass loss prescriptions. %However, theoretical modelling of stellar wind mass loss for massive stars remains uncertain.\cite{1994A&AS..103...97M,2014ARA&A..52..487S,2017A&A...603A.118R,2020ApJ...890..113B,2021arXiv210908164V}
% and by other stellar model parameters such as, e.g., overshooting \cite{2021MNRAS.504..146V}. \ssb{removed this comment, else we need to show this model uncertainty as well. I run the model and this would lead to even more massive BHs.}

The stellar winds of massive stars, in addition to determining the final stellar mass, also affect stellar structure and the star's position in the Hertzsprung--Russell (HR) diagram.\cite{1986ARA&A..24..329C, 1987A&A...182..243M} 
At solar metallicity, stellar winds of massive stars with zero-age main sequence (ZAMS) masses above $50\,M_\odot$ are so strong that they remove the hydrogen envelopes, exposing the cores of the stars. 
Unlike lower mass stars, these massive stars never expand to supergiant radii of $\sim 1000\,R_\odot$. Consequently, the feedback of wind mass loss onto stellar evolution makes the stellar tracks in the HR diagram turn to the blue before they become red supergiant stars. 
This established result of stellar evolution theory has been overlooked in models for the formation of merging BBHs. These models most often employ rapid binary population-synthesis (BPS) codes (e.g., \texttt{BSE},\cite{2002MNRAS.329..897H} \texttt{COSMIC},\cite{2020ApJ...898...71B} \texttt{COMPAS},\cite{2022ApJS..258...34R} \texttt{binary\_c},\cite{2004MNRAS.350..407I,2006A&A...460..565I,2009A&A...508.1359I}  \texttt{MOBSE},\cite{2018MNRAS.474.2959G} and \texttt{StarTrack}\cite{2002ApJ...572..407B}), where the impact of mass loss, either due to stellar winds or binary interactions, on the stellar properties is not treated self-consistently.
%In the light of the empirical evidence that the most massive BH ever observed in the Milky Way, the BH accretor in Cygnus X-1 has a mass $\sim 21\,M_\odot$,\cite{2021Sci...371.1046M} binary population synthesis (BPS) models of isolated binary evolution at solar metallicities have been directly or indirectly tuned to predict a maximum BH mass at $Z_\odot$ of $\sim 20\,M_\odot$.\cite{2010ApJ...714.1217B} Consequently, BPS models do not predict the existence of merging BBHs with masses $\gtrsim 20\,M_\odot$\cite{2016Natur.534..512B} at solar metallicity. Most of these models employ rapid BPS techniques which do not solve the stellar structure equations of the stars self-consistently along with the orbital evolution of the binary (see, e.g., \texttt{BSE},\cite{2002MNRAS.329..897H} \texttt{COSMIC},\cite{2020ApJ...898...71B} \texttt{COMPAS},\cite{2022ApJS..258...34R} \texttt{binary\_c},\cite{2004MNRAS.350..407I,2006A&A...460..565I,2009A&A...508.1359I}  \texttt{MOBSE},\cite{2018MNRAS.474.2959G} and \texttt{StarTrack}\cite{2002ApJ...572..407B}). 
%In addition, to rapidly simulate the evolution of single stars, 
Instead, the aforementioned BPS models employ fitting formulae to precalculated single stellar evolutionary tracks, up to ZAMS masses of $50\,M_\odot$, at constant mass (\texttt{SSE} code\cite{2000MNRAS.315..543H}).\cite{1998MNRAS.298..525P} 
To model the evolution of the progenitors of stellar-mass BHs, these rapid BPS models extrapolate \texttt{SSE} stellar tracks up to $150\,M_\odot$. 
Mass loss due to stellar winds is applied a posteriori without accounting for the discussed mass loss feedback onto the stellar evolution. 
Rather, mass loss is implemented by interpolating a new stellar evolutionary track with the mass decreased by the mass lost through stellar winds in the given time-step, conserving the information of the stellar core of the original track.\cite{2000MNRAS.315..543H} 
The limitations of \texttt{SSE} stellar tracks in predicting the maximal radial expansion of massive stars at solar metallicity has been known since its conception.\cite{2000MNRAS.315..543H} 
However, less well studied is its impact on massive binary evolution and its implications for merging BBH formation, 
making it a major uncertainty in our current understanding.\cite{2020MNRAS.497.4549A,2022arXiv221115800R}
%making it a potential caveat for many studies in the literature. 
Although rapid BPS has been an invaluable tool in predicting and interpreting BBH and XRB observations thus far, advances in the simulations of binary stars\cite{2015ApJS..220...15P,2022arXiv220205892F} allow us to revisit the topic.

%a key limitation of them lies in the precalculated stellar evolution formulae they apply, that are based on constant-mass models of ZAMS masses $\leq 50\,M_\odot$.
%The key underlying assumption made in rapid BPS studies leveraging the \texttt{SSE} stellar tracks is that the process of wind mass loss does not alter the stellar evolution.
Here, we use the \texttt{POSYDON}\cite{2022arXiv220205892F} framework which is a novel  BPS that models self-consistently both stars and their binary interactions simultaneously. Other than accounting for the feedback of stellar wind mass loss onto the stellar evolution, \texttt{POSYDON} binary stellar models keep track in a detailed way the angular momentum transport within the stellar interiors of the binary components, and between the two binary components and the orbit. Additionally, \texttt{POSYDON} include self-consistent calculations of binary interactions and mass-transfer phases. To achieves this level of model sophistication, \texttt{POSYDON} uses pre-computed grids of detailed stellar structure and binary evolution simulations, performed with the \texttt{MESA}\cite{2011ApJS..192....3P,2013ApJS..208....4P,2015ApJS..220...15P,2018ApJS..234...34P,2019ApJS..243...10P} code, while at the same time maintaining the flexibility of rapid BPS codes.
Other rapid BPS codes that rely on detailed models and, hence, not on \texttt{SSE} stellar tracks, do exist. 
For example, \texttt{ComBinE}\cite{2018MNRAS.481.1908K} and \texttt{SEVEN}\cite{2015MNRAS.451.4086S} use look-up tables for the properties of single stars, based on grids of precalculated detailed, single-star models that take into account wind mass-loss self-consistently, while treating binary interactions using approximate prescriptions and parametrizations similar to the mentioned rapid BPS codes. 
Alternatively, \texttt{BPASS}\cite{2017PASA...34...58E} uses extensive grids of detailed binary evolution models computed with a custom version of the Cambridge \texttt{STARS} binary evolution code,\cite{2009MNRAS.396.1699S} where both stars are followed in detail, but only one at a time for computational constraints. However, none of the aforementioned BPS codes, other than \texttt{POSYDON}, employ fully self-consistent detailed stellar-structure and binary simulations, including angular momentum transport in stellar interiors, which are essential to achieve accurate predictions of double compact object properties.

In this work, we examine how self-consistent modelling of mass loss, stellar structure, and binary interactions impacts the formation of merging BBHs at solar metallicity. 
First, we discuss BH formation in the context of single stellar evolution, focusing on the key aspects that determine the maximum BH mass. 
%Massive stars at solar metallicity ($\gtrsim 50\,M_\odot$), in contrast to their lower mass counterparts, expand less during their evolution, and do not reach supergiant radii $\sim 1000\,R_\odot$, while at the same time form BHs of up to $\gtrsim 30 \, M_\odot$.
Second, we model the observational properties of merging BBHs in Milky-Way-like galaxies. 
In contrast to previous studies,\cite{2013ApJ...779...72D,2019MNRAS.490.3740N,2022MNRAS.516.5737B} our model predicts a non-negligible BBH merger rate at solar metallicity with BHs as massive as $\sim 30\,M_\odot$.
%In contrast to previous predictions based on rapid BPS, we find that BHs can have masses as high as $~30\,M_\odot$ a non-negligible rate of merging BBHs at solar metallicity
%Finally, we comment on other observable implications of the existence of such massive BHs at solar metallicity in the context of GW detections, XRBs and astrometric binaries. 

 \begin{figure}[H]
\centering
\includegraphics[]{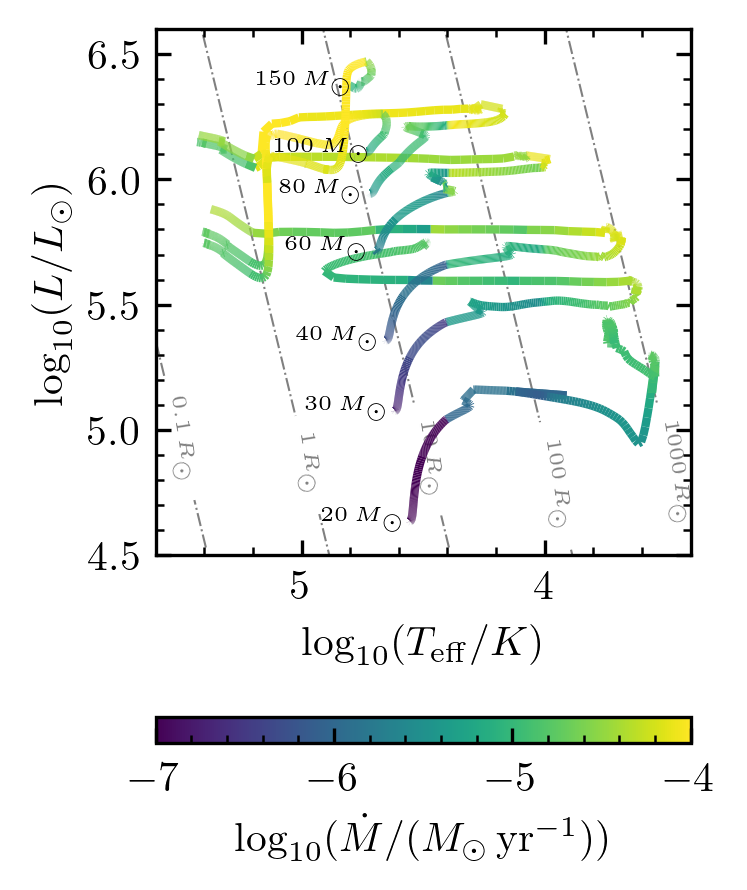}
\includegraphics[]{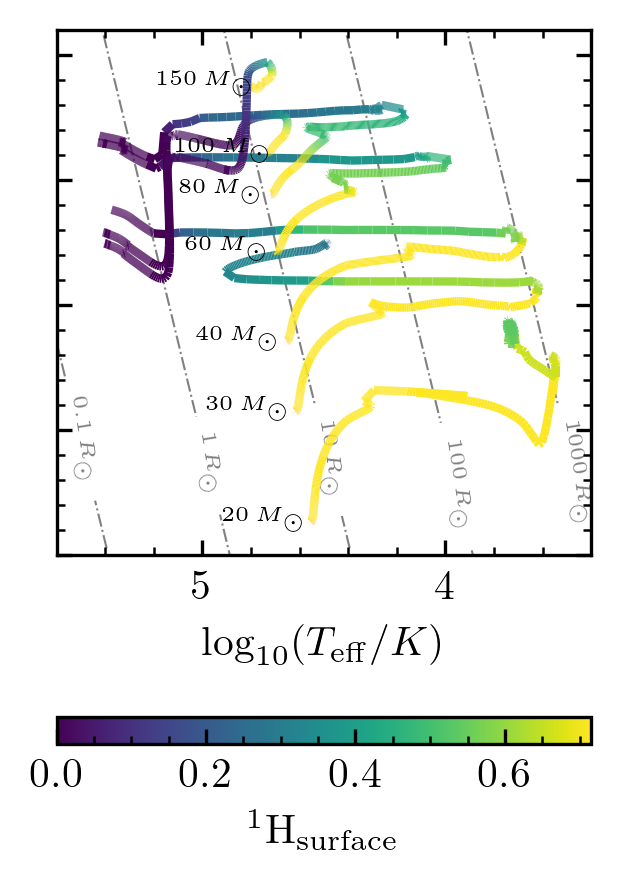}
\includegraphics[]{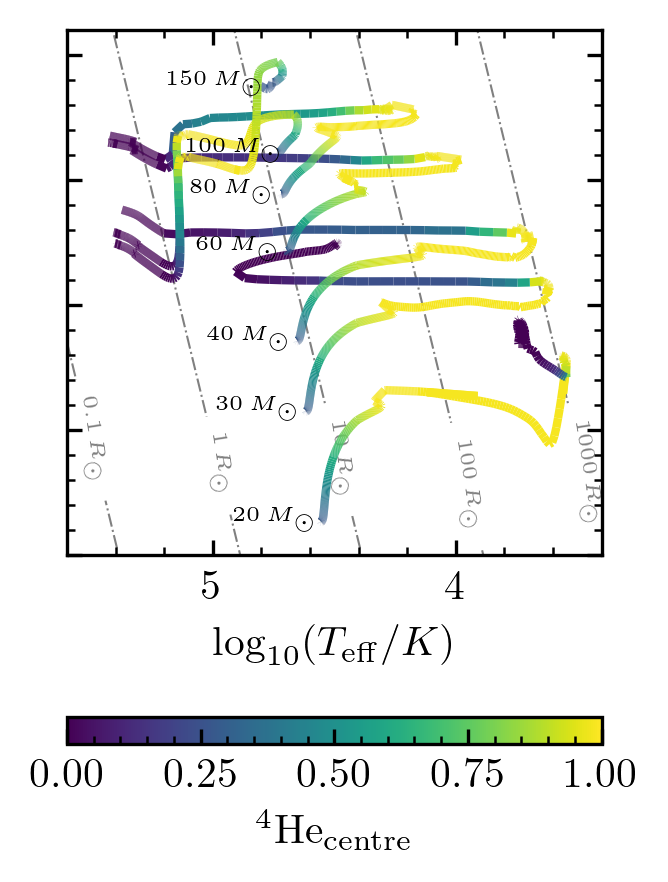}
\caption{Hertzsprung--Russell diagram for stars with masses between $20\,M_\odot$ and $150\,M_\odot$. 
Stellar tracks are displayed up to central carbon depletion. The instantaneous mass loss (\textit{left}), surface hydrogen abundance (\textit{middle}) and centre helium abundance (\textit{right}) are indicated by the track's colour. 
Stars with masses $\gtrsim 80\,M_\odot$ reach the Wolf--Rayet phase while still burning hydrogen in their cores.}
\label{fig:default_model}
\end{figure}

\begin{figure}[H]
\centering
\includegraphics[]{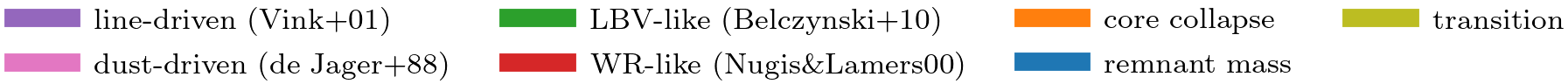}
\includegraphics[]{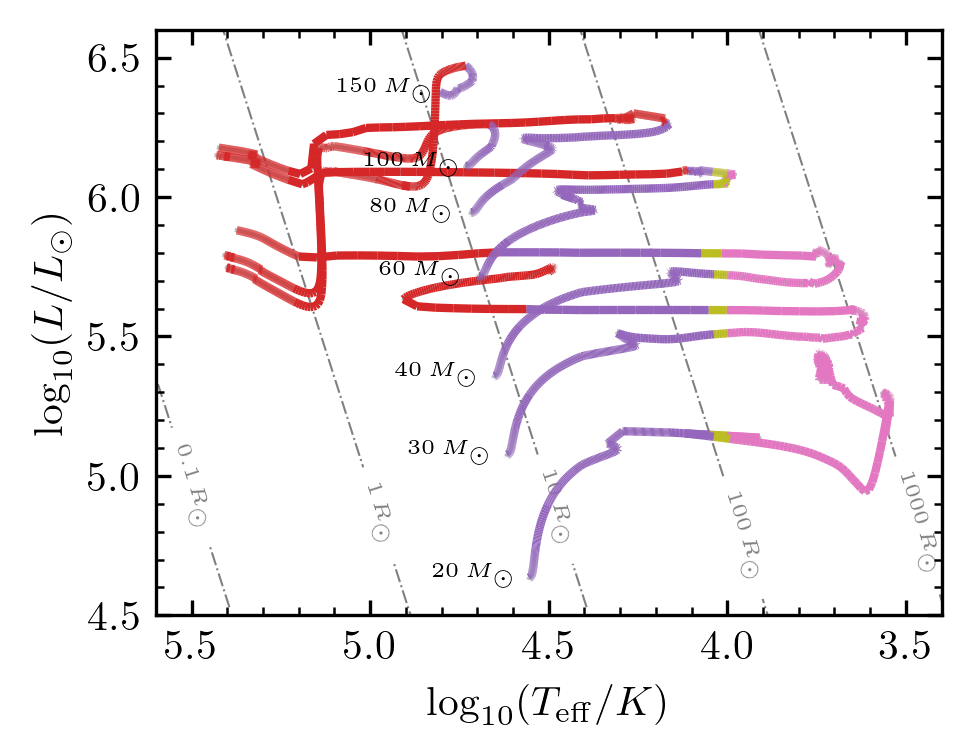}
\includegraphics[]{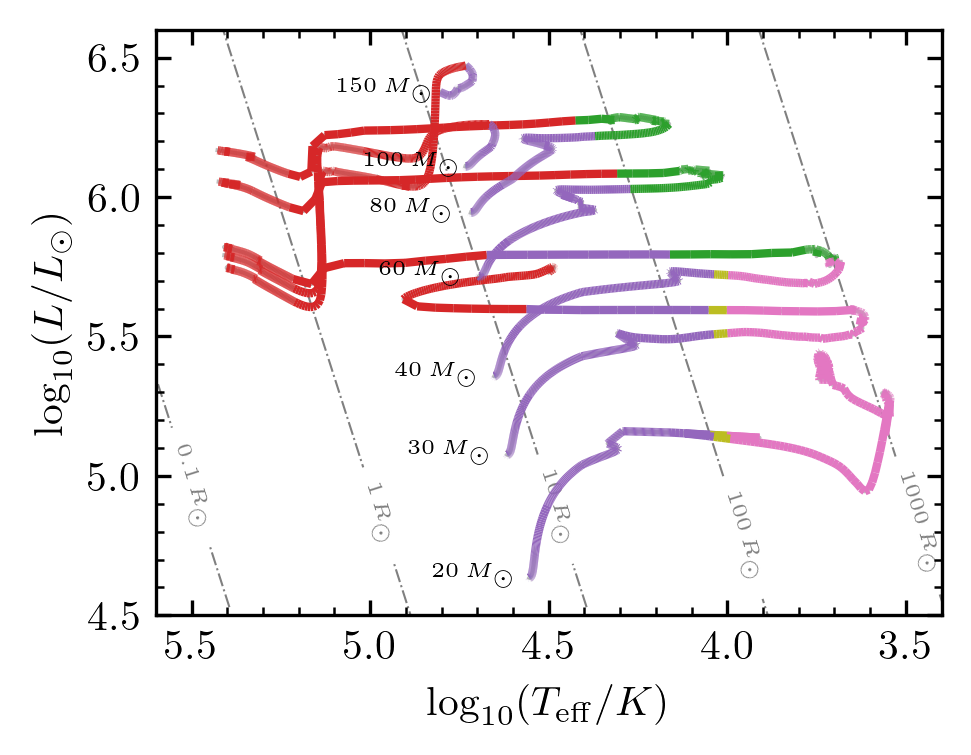}
\includegraphics[]{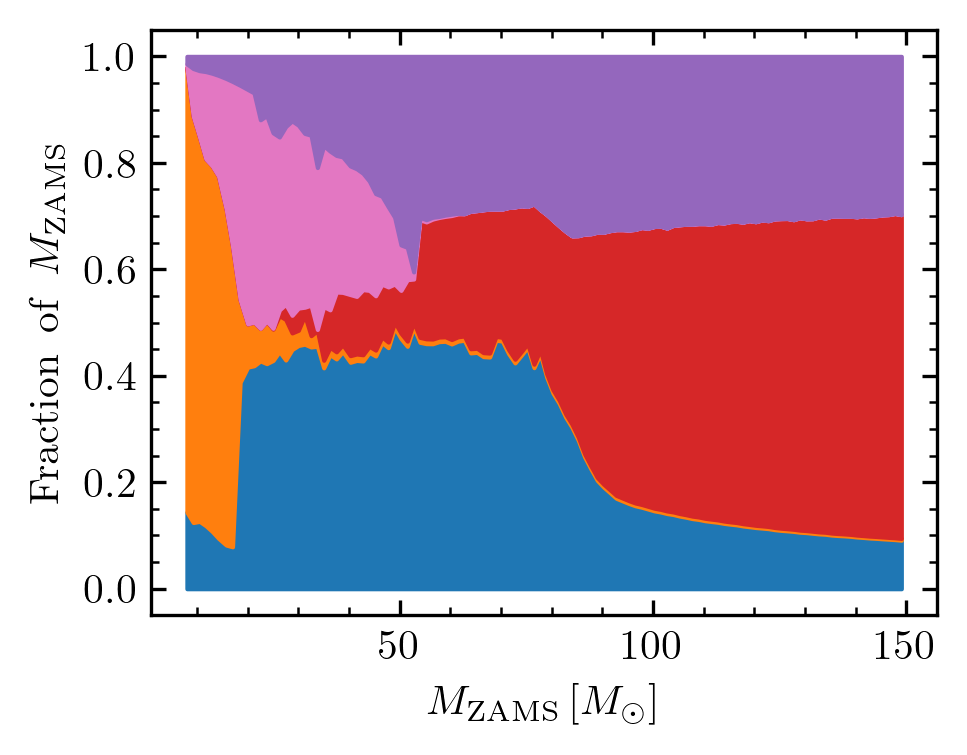}
\includegraphics[]{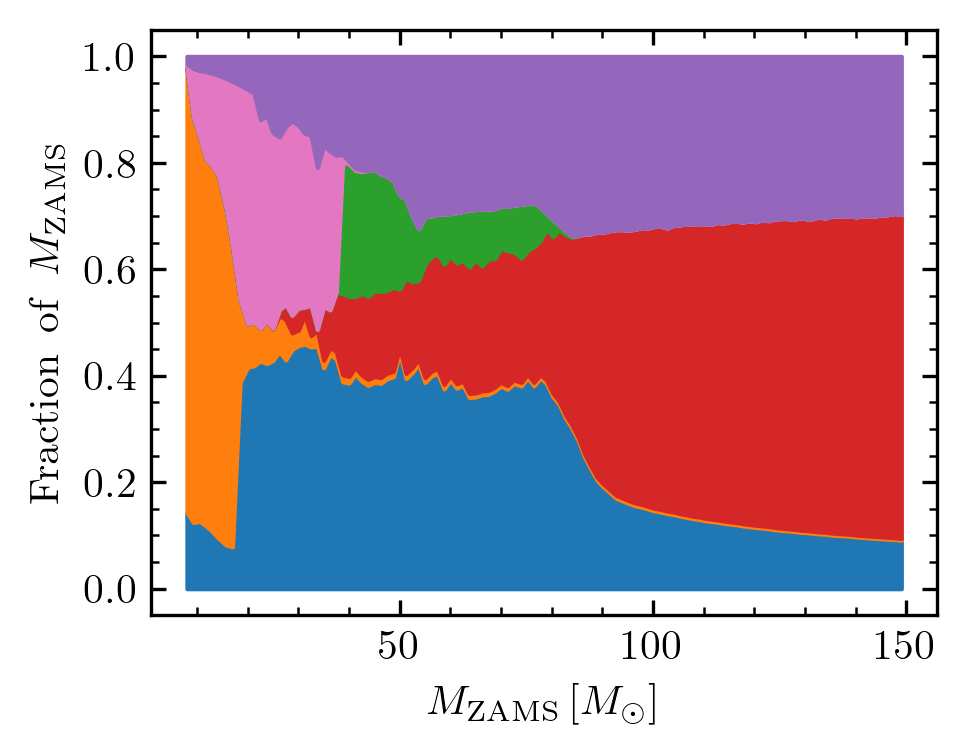}
\caption{(\textit{Top}) Hertzsprung--Russell diagram for stars with masses between $20\,M_\odot$ and $150\,M_\odot$. 
We denote with different colours indicating the different stellar wind prescriptions implemented in our default model (\textit{top-left}) and model variation that includes LBV-like winds (\textit{top-right}). 
(\textit{Bottom}) Fraction of mass lost due to different mass loss mechanisms during the entire life of a \texttt{POSYDON} single star at solar metallicity as a function of ZAMS mass in the mass range between $8\,M_\odot$ and $150\,M_\odot$ sampled every $1.25\,M_\odot$. 
In our model, BH formation for single stars occurs for $M_\mathrm{ZAMS} \gtrsim 17.3 \, M_\odot$.}
\label{fig:winds}
\end{figure}

\section*{Results}

\subsection*{The maximum BH mass from single stellar evolution at solar metallicity}

We first present a subsample of the high-mass, single-star models at solar metallicity, and discuss their phenomenological evolution in the HR diagram. 
In Figure~\ref{fig:default_model}, we show stellar tracks in the mass range from $20\,M_\odot$ to $150\,M_\odot$, where we use a colour map to indicate the stellar-wind mass loss, as well as the surface ${}^1\mathrm{H}$ and centre ${}^4\mathrm{He}$ abundances of the stars during their evolution. 
Stars in this mass range are progenitors of stellar-mass BHs. 
In agreement with other state-of-the-art single-star models like \texttt{GENEVA}\cite{1987A&A...182..243M,1994A&AS..103...97M,2003A&A...404..975M,2012A&A...537A.146E} and \texttt{MIST},\cite{2016ApJ...823..102C} our stellar tracks of massive stars ($M_\mathrm{ZAMS} \gtrsim 50 \, M_\odot$) showed in Figure~\ref{fig:default_model} 
%, stellar winds are sufficiently strong to remove the hydrogen envelope and reveal the inner layers of the star that have been process by nuclear burning, and alter the stellar surface composition of the stars soon after they leave the main sequence, which consequently alter the subsequent stellar evolution.This process, makes the stars 
expand less than those of lower-mass stars, after they leave the main sequence, and never reach the red supergiant phase where stars have radii of $\sim 1000 \, R_\odot$.
Stellar winds of the most massive stars ($\gtrsim 80 \, M_\odot$) during the main sequence are so strong that deplete the hydrogen envelopes and induce the Wolf--Rayet (WR) phase while the stars are still burning hydrogen in their cores. 
%In the central and right panels of Figure~\ref{fig:default_model}, we use a colour map to show the surface ${}^1H$ abundance and the centre ${}^4\mathrm{He}$ of the single star along their evolution in the HR diagram. 
Indeed, Figure~\ref{fig:default_model} shows that the most massive stars reach the WR stage, i.e. when the surface ${}^1\mathrm{H}$ fraction drops below $0.4$ (see Methods), before the centre ${}^4\mathrm{He}$ fraction becomes $1$.
Hence, these stars will experience a longer-lived WR phase and, consequently, experience most of their mass loss due to WR stellar winds. These stellar tracks result in final masses smaller than those of lower-mass stellar models. 
This distinct radial expansion signature of massive single stellar evolution at solar metallicity is not present when we consider a model variation without stellar-wind mass loss as the lack of such feedback onto stellar evolution would cause the stars to expand to supergiant radii independently of their ZAMS mass (see Supplementary information).

Complementarily, the top-left panel of Figure~\ref{fig:winds} shows the different stellar-wind mass loss prescriptions applied to the tracks according to the fiducial (default) model of \texttt{POSYDON} (see Methods). 
\texttt{POSYDON} \texttt{v1.0} default stellar models do not account for any luminous blue variable (LBV)-type winds. 
This choice was made given the uncertain estimates of LBV-type wind mass losses.\cite{2017RSPTA.37560268S} 
Regardless, here, we also consider a model variation of our default stellar assumptions which accounts for LBV-type winds when stars evolve above the Humphreys--Davidson limit (see Methods for further details). For comparison, the alternative model including LBV-type winds is shown in the top-right panel of Figure~\ref{fig:winds}.

To illustrate how much mass is lost by the stars in the different stellar-wind regimes, in the bottom panels of Figure~\ref{fig:winds}, we show the fraction of mass lost due to a given stellar wind prescription or core collapse with respect to the initial ZAMS mass. 
%To generate this figure, we include stellar tracks in the mass range from $8\,M_\odot$ to $150\,M_\odot$ with a resolution of $1.25\,M_\odot$. 
In both considered models, stars with $M_\mathrm{ZAMS}\gtrsim 17.3\,M_\odot$ may form a BH, given the assumed Patton\&Sukhbold20\cite{2020MNRAS.499.2803P} core-collapse prescription, where we also account for up to $0.5\,M_\odot$ mass loss due to neutrinos in the core-collapse.\cite{2022arXiv220205892F} 
In contrast, less massive stars explode into a supernova to form a neutron star where a large fraction of the ZAMS mass is ejected in the process. The bottom panels of Figure~\ref{fig:winds} show that in both models, massive stars with $M_\mathrm{ZAMS} \gtrsim 50\,M_\odot$ lose most of their mass through WR winds. 

\begin{figure}[ht]
\centering
\includegraphics[]{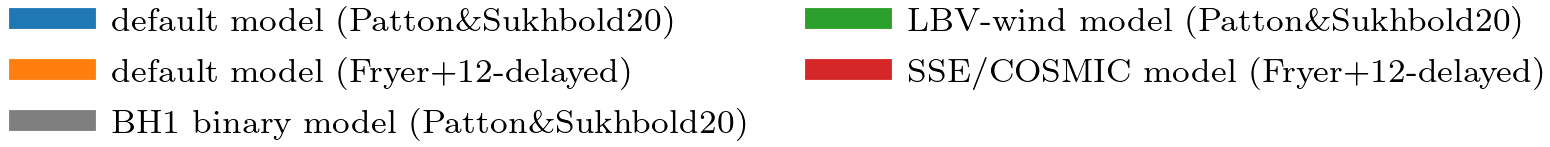}
\includegraphics[]{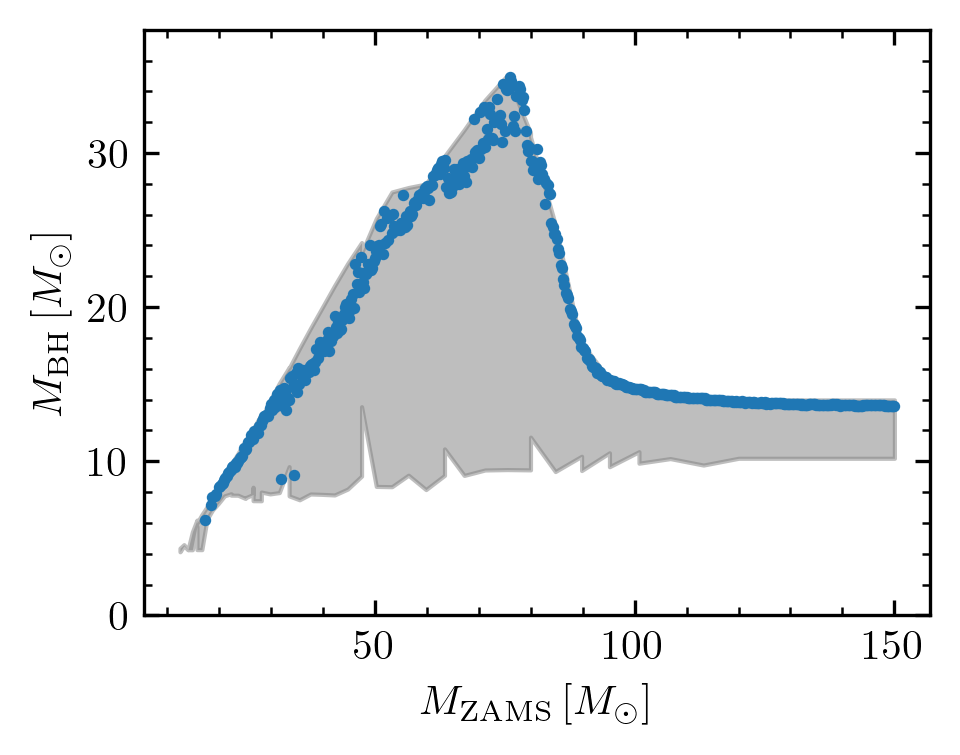}
\includegraphics[]{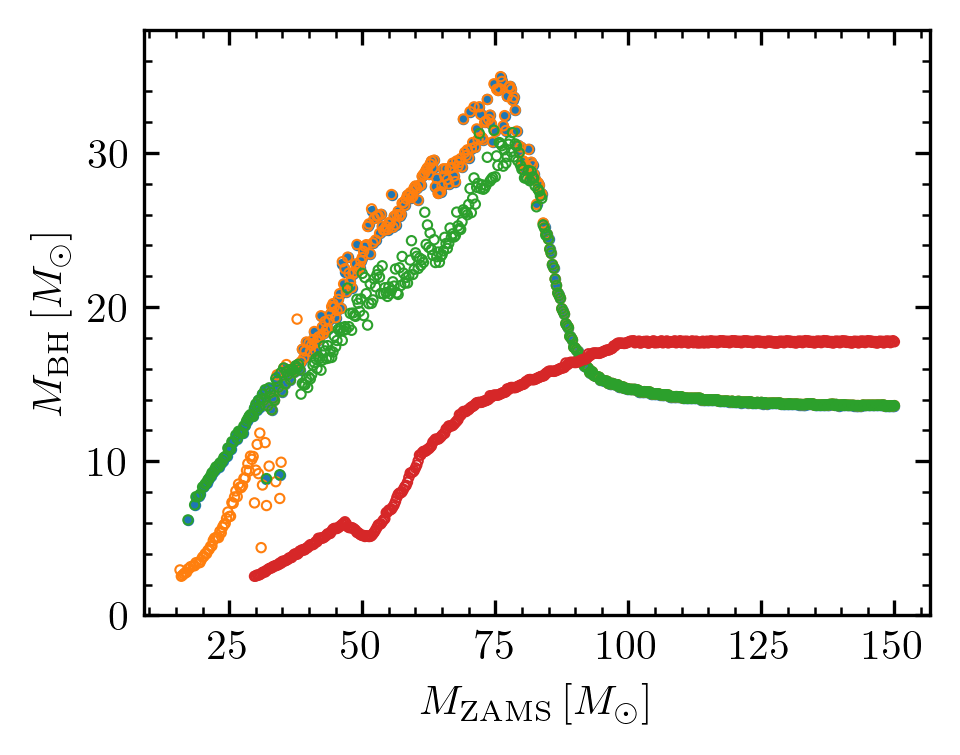}
\caption{(\textit{Left}) BH mass from single stellar evolution for a given ZAMS stellar mass at solar metallicity (markers) and the first-born BH mass of binary systems (gray area), with the extend of the grey area caused by the diversity of binary interactions starting at different orbital periods covered in {\tt POSYDON}'s binary grids. (\textit{Right}) We show the BH mass, resulting from single stars, given two different core-collapse assumptions for our default model according to the legend and the model variation including LBV-like winds. 
For comparison, we show BH masses as predicted by the \texttt{SSE} stellar models as implemented in \texttt{COSMIC}\cite{2020ApJ...898...71B} given one of the considered core-collapse prescriptions.}
\label{fig:BH_mass}
\end{figure}

To visualise the effect of wind mass loss of single stars onto the BH mass, in Figure~\ref{fig:BH_mass}, we show the BH mass as a function of ZAMS mass.
The left panel of Figure~\ref{fig:BH_mass}, shows a nearly monotonic increasing relation between the stars ZAMS masses and the BH masses up to $M_\mathrm{ZAMS} \simeq 80\,M_\odot$. 
The maximum BH mass predicted by our default model is $\simeq 35.0 \, M_\odot$. 
In the same figure, for comparison, we also show the BH mass range corresponding to the formation of the first-born BH in the grid of binary-star models of \texttt{POSYDON}. 
The difference in the BH mass spectrum for a given ZAMS mass is caused by mass loss during binary mass transfer, which carries away some of the stellar mass prior to their core-collapse. 

The right panel of Figure~\ref{fig:BH_mass} shows the model variation accounting for LBV-type enhanced wind mass loss. 
This alternative model leads to slightly less massive BHs at solar metallicity with a maximum BH mass of $\simeq 31.5\,M_\odot$. 
In the same figure, we also show a different assumption for the core-collapse, the Fryer+12-delayed\cite{2012ApJ...749...91F} prescription, which is commonly used in rapid BPS studies. 
Both assumed core-collapse prescriptions predict direct collapse for stars with carbon--oxygen cores above $11\,M_\odot$ at carbon depletion. 
Hence, in practice, both prescriptions predict the same BH masses for stars with $M_\mathrm{ZAMS}\gtrsim 30\,M_\odot$, while Fryer+12-delayed prescription predicts mass ejection for stars with $M_\mathrm{ZAMS}\in [15.8,30]\,M_\odot$, and, hence, less massive BHs. 
We also compare our model predictions with \texttt{SSE} stellar models. 
In the right panel of Figure~\ref{fig:BH_mass}, we show the BH masses as predicted by the Fryer+12-delayed prescription applied to \texttt{SSE} stellar models. 
In contrast to the \texttt{POSYDON} stellar models, \texttt{SSE} predicts much smaller BH masses for stars with $M_\mathrm{ZAMS}\lesssim 90\,M_\odot$ and a maximum BH mass of $\simeq 17.8\,M_\odot$ for $M_\mathrm{ZAMS}\gtrsim 90\,M_\odot$, noticeably smaller than the maximum BH mass predicted by \texttt{POSYDON}.
The source of the difference originates from the overestimation of the radial expansion in \texttt{SSE} stellar tracks, which in turn shifts the fraction of the time that each track spends in the different stellar-wind regimes. 
In our models, we find that most of the stellar mass is lost by WR winds. 
By remaining more compact during their evolution, the BH progenitor stars avoid by the most part the regions of the HR diagram where dust-driven or LBV-type winds are expected to efficiently drive high mass-loss rates. 
In contrast, \texttt{SSE} predicts that massive stars ($M_\texttt{ZAMS} \gtrsim 50\,M_\odot$) spend a considerable fraction of their evolution in the LBV-wind regime and lose most of their mass due to LBV-like winds (see the left panel of Figure 9 in Dorozsmai\&Toonen22\cite{2022arXiv220708837D} which can be directly compared to our bottom panels of Figure~\ref{fig:winds}). Correctly modeling the the maximal expansion of massive stars is crucial to accurately model merging BBHs and interpret GW observations.
%The lack of the self-consistent treatment of wind mass loss leads to the \texttt{SSE} tracks to over-expanding up to $\simeq 1000 \,R_\odot$ and spend much more time in the LBV-wind regime, over-predicting the mass loss during this phase. When accounting for LBV-like winds in our models, bottom-right panel of Figure~\ref{fig:BH_mass_LVC}, we still find that the mass loss of massive stars is dominated by Wolf-Rayet winds. %Given the assumed mass loss prescription, we find that stars undergoing LBV winds in the mass range $M_\mathrm{ZAMS} \in [40,80]\,M_\odot$ can experience up to $20\%$ grater mass loss. In the default model the maximum BH mass at solar metallicity is $\simeq 35\,M_\odot$, while for the LBV-wind model is $\simeq 31.5\,M_\odot$.

\begin{figure}[ht]
\centering
\includegraphics[]{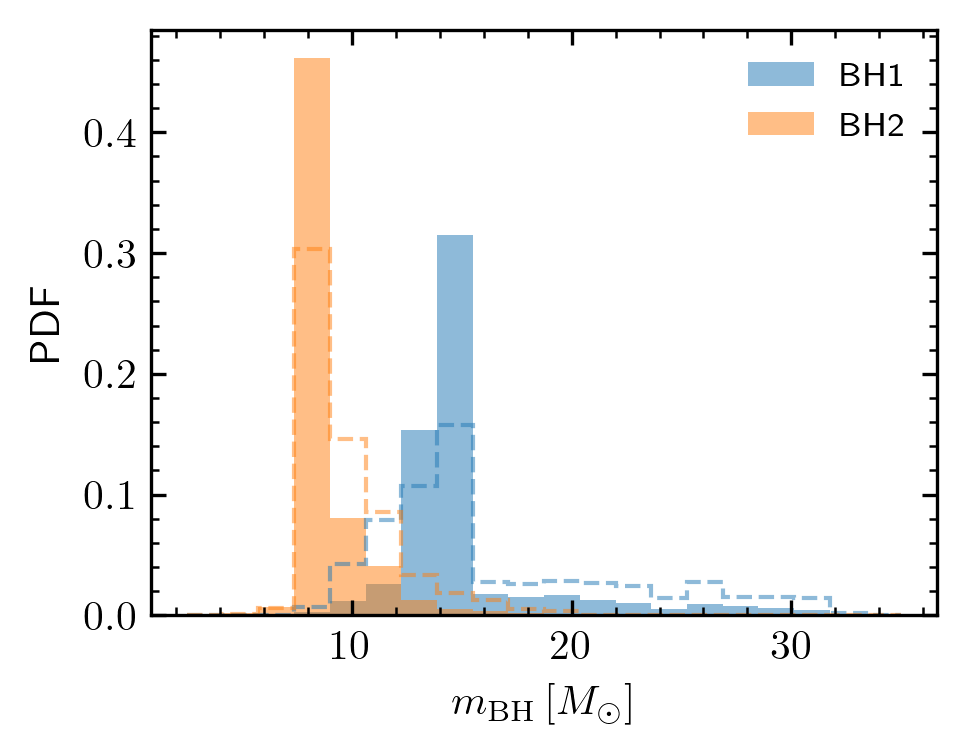}
\includegraphics[]{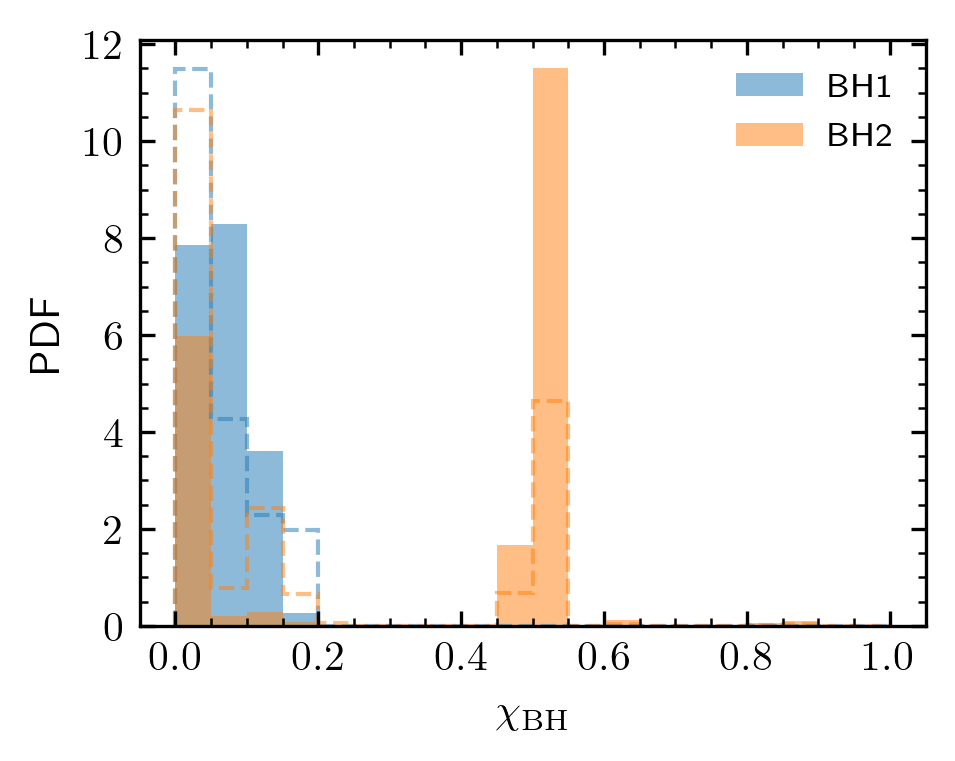}
\caption{Merging binary black hole population at solar metallicity. 
Filled bins indicate the underlying (intrinsic) BBH distribution, while dashed lines illustrate selection effects for a ground-based GW detector. Here, we assume Advanced LIGO\cite{2015CQGra..32g4001L} at design sensitivity as an example.
We distinguish with the indices ${1,2}$ the first- and second-born BHs, respectively.  (\textit{Left}) BBH component mass distributions of merging BBHs formed from binary stars at solar metallicity. 
(\textit{Right}) BBH component spin-magnitude distributions of merging BBHs formed from binary stars at solar metallicity. 
Systems with $\chi_\mathrm{BH} > 0.4$ are believed to be associated with a luminous LGRB at the time of BH formation.\cite{2022A&A...657L...8B}
}
\label{fig:BH_mass_LVC}
\end{figure}

\subsection*{The population of merging BBHs at solar metallicity}

% LIGO-Virgo
The existence of massive BHs ($\simeq 30\,M_\odot$) from single stars alone at solar metallicity does not directly imply that of similarly massive merging BBHs at solar metallicity. 
Compared to the progenitors of lower mass BHs, the progenitor stars of these massive BHs remain more compact during their evolution and avoid mass transfer before the formation of the first-born BH. For binary stars with primary ZAMS masses below $50\,M_\odot$, Roche-lobe overflow occurs for systems with initial orbital periods of $\lesssim 10^{3.5}~\mathrm{days}$, while for primaries with masses from $50 \,M_\odot$ to $80 \,M_\odot$ only for initial periods below the range $10^{1.8}$--$10^{2.5}~\mathrm{days}$ depending on the binary mass ratio. 
Stars with birth masses of more than $80 \,M_\odot$ at solar metallicity can avoid mass transfer for initial periods down to a few days depending on the mass ratio (see Figure~\ref{fig:grids} in Supplementary information).
Our \texttt{POSYDON} BPS model predicts the formation of merging BBH from both: (i) binaries that undergo a stable mass transfer prior to the formation of the first-born BH, and (ii) more massive binaries that avoid such phase. 
After the formation of the first-born BH, depending on the secondary mass, these systems will undergo a reverse stable mass-transfer\cite{2017MNRAS.471.4256V,2017MNRAS.465.2092P} or a common-envelope phase.\cite{2007PhR...442...75K,2016Natur.534..512B}
In our BPS model at solar metallicity, all systems experiencing a common-envelope prior to the formation of the second-born BH avoid undergoing any mass transfer before the formation of the first-born BH.\footnote{In our results, we exclude binaries that experience reverse mass-transfer prior to the formation of the first-born BH. This is due to these systems not being treated properly by {\tt MESA} in the version used in \texttt{POSYDON} \texttt{v1.0}. This problem affects approximately 4\% of tracks in the grid of binaries consisting of two hydrogen-rich main-sequence stars (see Section 5.5 of the {\tt POSYDON} instrument paper\cite{2022arXiv220205892F}).}
In contrast, only one fourth of the systems undergoing a stable mass transfer prior to the formation of the second-born BH avoid undergoing any mass transfer before to the formation of the first-born BH.
These alternative evolutionary pathways that avoid a stable mass-transfer episode before the formation of the first-born BH stand in contrast to what commonly found with rapid BPS models at sub-solar metallicity where such phase is present.\cite{2020A&A...635A..97B,2021A&A...647A.153B} 

In the left panel of Figure~\ref{fig:BH_mass_LVC}, we show the component BH mass distribution of the intrinsic merging BBH population at solar metallicity. 
The distribution is constructed by distributing the synthetic BBH population obtained from the BPS model following the star formation history of the Universe consistent with the stars having solar metallicity. 
Specifically, to obtain the underlying merging BBH populaiton at solar metallicity, we use the \texttt{IllustrisTNG}\cite{2015A&C....13...12N} simulated star formation rate happening in the metallicity range $[0.5Z_\odot,2Z_\odot]$. 
For comparison, in Figure~\ref{fig:BH_mass_LVC} we also indicate a similar distribution that accounts for selection effects of a ground-based GW detector. As an example, we consider Advanced LIGO\cite{2015CQGra..32g4001L} at design sensitivity\cite{2018LRR....21....3A} (see the Methods section for further details). 
We find that the BH mass spectrum of merging BBHs at solar metallicity spans the same range of BH masses formed from single stars. 
However, we find fewer BHs above $16\,M_\odot$. 
We quantify the fraction of first-born BHs with masses above $20 \,M_\odot$ to be $f_{m_\mathrm{BH1} > 20\,M_\odot} =0.10$, while those of above $30 \,M_\odot$ constitute a fraction of $f_{m_\mathrm{BH1} > 30\,M_\odot} = 0.01$. 
When accounting for LIGO selection effects, these fractions increase to $f^\mathrm{det}_{m_\mathrm{BH1} > 20\,M_\odot}=0.24$ and $f^\mathrm{det}_{m_\mathrm{BH1} > 30\,M_\odot}=0.03$, respectively. 
These results cannot be compared directly to the observed sample of BBHs from the LIGO Scientific, Virgo and KAGRA Collaboration,\cite{2019PhRvX...9c1040A,2021PhRvX..11b1053A,2021arXiv210801045T,2021arXiv211103606T} as observed merging BBHs come from a range of metallicities. One would expect that accounting for the lower metallicity stellar populations would increase the average BH mass, in which case the aforementioned fractions can be interpreted as lower limits when compared to the total observed population

Furthermore, in contrast to past studies,\cite{2013ApJ...779...72D,2019MNRAS.490.3740N,2022MNRAS.516.5737B} we find a non-negligible BBH merger rate efficiency per stellar mass formation of $4.4\times 10^{-7}\,M_\odot^{-1}$ at solar metallicity. 
Of the total rate, around $40\%$ comes from systems undergoing a common-envelope phase, around $45\%$ undergo a stable mass-transfer phase before the formation of each BH, and around $15\%$ experience a stable mass-transfer phase only after the formation of the first-born BH. 
Accounting again only for the Universe star formation rate in the metallicity range $[0.5Z_\odot,2Z_\odot]$, we can estimate the intrinsic local merger rate density of BBHs formed at solar metallicity to $8.6 \, \mathrm{Gpc}^{-3}\mathrm{yr}^{-1}$ and the corresponding detection rate for Advanced LIGO at design sensitivity to around $25 \, \mathrm{yr}^{-1}$. 
Again, these rates are not to be directly compared to the ones inferred from the LIGO Scientific, Virgo and KAGRA Collaboration\cite{2021arXiv211103634T} due to the omission in our models of the BBH population formed at low-metallicity environments. Nevertheless, should our predicted rates be higher than the observed ones, our model would be inconsistent with observations. 
%Of the total intrinsic local BBH merger rate density, only $30\%$ comes from systems undergoing a common envelope event.

%In contrast to other studies employing rapid BPS leveraging the \texttt{SSE} stellar models \ssb{cite}, we find that binaries with massive primary stars, $M_\mathrm{ZAMS} \gtrsim 80 M_\odot$ do not undergo mass transfer before the formation of the first-born BH. This discrepancy rises from the fact that, in our models, these massive stars expand to much smaller maximal radii (cf. Figure~\ref{fig:wind_mdot}).

%We estimate  These rates should be carefully interpreted as they are caveat to the fact that the star formation rate of the Universe does not occur entirely at $Z_\odot$. Hence, it has the sole purpose of verifying that the model prediction does not over-predict the current observational constraints of GWTC-3 of $23.9^{+14.9}_{-8.6} \, \mathrm{Gpc}^{-3}\mathrm{yr}^{-1}$.\cite{2021arXiv211103634T}

% LGRBs
In the right panel of Figure~\ref{fig:BH_mass_LVC}, we show the component BH dimensionless spin parameters of the modeled merging BBH population at solar metallicity.
Our BPS model predicts an underlying fraction of $19\%$ of merging BBHs with first-born BH spins in the range $ 0.1 \leq \chi_\mathrm{BH1} \leq  0.2$. 
The progenitor stars of these mildly spinning BHs have masses $M_\mathrm{ZAMS} > 50 \, M_\odot$ and orbits with initial periods of a few days. 
They may undergo episodes of Case-A Roche-lobe overflow mass-transfer or stable contact phase, which limits the expansion of the orbit, or in some cases avoid mass-transfer all together. 
In all cases, the two stars remain sufficiently close to each other where tides can maintain the primary star rotating until the formation of a mildly-spinning first-born BHs when efficient angular momentum transport is assumed (see Figure~\ref{fig:grids} in Supplementary information).
%Because these stars expand less during their evolution compared to lower mass stars before the formation of the first-born BH, they experience less specific angular momentum loss, maintaining the rotation gained . 
This result is opposed to the common assumption in recent rapid BPS studies\cite{2015ApJ...800...17F,2018A&A...616A..28Q,2020A&A...635A..97B,2021A&A...647A.153B} that efficient angular momentum transport leads to the formation of first-born BHs with negligible spin.
%\footnote{\ssb{point out the prediction of Belckzinski from single stars? Decided not to, since those results comes from assumptions on single stellar models run at arbitrary initial high rotation}}
%\ssb{we decided to omit this channel} There is another smaller sub-population of first-born BHs with high spins ($\chi_\mathrm{BH1} \in [0.4,0.6]$) \ssb{quantify}. These highly spinning BHs are formed from systems undergoing double common envelope, which leads to the formation of two Wolf-Rayet stars in close orbits which experience tidal spin-up and forming two highly rotating BHs.\cite{2017ApJ...842..111H,2021ApJ...921L...2O} 

The spin of the second-born BH is determined by the tidal interactions during the BH--WR phase. 
In agrement with past studies, we find that only systems evolving through a common envelope reach a close enough orbital separation to experience significant tidal spin-up.\cite{2021RNAAS...5..127B}
The formation of BHs with high spins of $\chi_\mathrm{BH} > 0.4$ has been associated with luminous LGRBs at the formation of the BH.\cite{2022A&A...657L...8B} 
Our prediction for the formation of BHs with $\chi_\mathrm{BH} > 0.4$ at solar metallicity and hence existence of LGRBs at solar metallicity (in contrast to earlier work by Bavera+22\cite{2022A&A...657L...8B} based partly on rapid BPS calculations) is consistent with the solar-like inferred metallicities of a small fraction of LGRB host galaxies.\cite{2019arXiv190402673G}

%\ssb{TODO: This is a placeholder for a paragraph that will do a quick comparison of our results with Combine, BPASS, and SVEN. Add a comment also to the new paper of Belczynski}
The results presented in this section are a direct consequence of the reduced maximal expansion of massive stars at solar metallicity. Previously mentioned rapid BPS codes based on detailed stellar models, i.e. not \texttt{SSE} stellar tracks, do share similar trends. For example,
\texttt{SEVN} models\cite{2015MNRAS.451.4086S,2019MNRAS.485..889S,2022arXiv221111774I} also predicts the formation of massive BHs at solar metallicity and a similar non-negligible BBH merger rate efficiency per solar mass. %shows similarly high mass BHs to be formed with a slightly higher rates than within our models. Those results are already present in their 2019 paper and not just in the recent one. At Z=0.02 they get BH masses up to 30Msun. They got a local merger rate of 90Gpc^{-3}yr^{-1}, which is an order of magnitude larger, which is integrated over redshift. They get about 1/10 in the number of mergers than usual for low metallicity, thus their rate with at Zsun only would be just a bit larger than the 8.6Gpc^{-3}yr^{-1}.
In contrast, \texttt{ComBinE} models\cite{2018MNRAS.481.1908K,2018PhDT.......146K}  predict a smaller maximal BH mass than the one predicted here. This is because these models assume stronger WR winds and less fall back mass during BH formation compared to our fiducial assumptions. 
%Given the reduced expansion of massive stars at solar metallicity, \texttt{ComBinE} BBH merger rate at solar metallicity are very sensitive to common envelope efficiency, which can cause a rate change by more than a factor of 100. 
%The default model just has a rate of 0.6Gpc^{-3}yr^{-1}, but the optimistic setup goes to 40Gpc^{-3}yr^{-1} for a bit more than half Zsun. The BH masses from ZAMS stars with masses above 150Msun go even below 10Msun again. But this is questionable, because here we are in the extrapolated regime.
%\texttt{METISSE}\cite{2020MNRAS.497.4549A,2022MNRAS.512.5717A} can create massive BHs at solar metallicity depending on the underlying stellar tracks. %They only show the initial to final mass relation at 0.1Zsun in their first paper. They didn't looked for BBH merger rates.
\texttt{BPASS} models\cite{2016MNRAS.462.3302E,2017PASA...34...58E} predict BH masses of a few solar masses grater than rapid BPS codes employing the \texttt{SSE} stellar tracks at solar metallicity but a lower BBH merger rates compared to the one presented here.
%At slightly lower metallicity the masses and rates become very similar. %Their masses go up to a bit more than 20Msun, while at Z=0.01 they go up to 40Msun BHs.
However, none of the aforementioned BPS models other than \texttt{POSYDON} employing fully self-consistent detailed stellar-structure and binary simulations can accurately predict merging BBH spin distributions, which have been shown to be crucial for differentiating BBHs formed via isolated binary evolution as opposed to other formation channels.\cite{2017Natur.548..426F}

\section*{Discussion \& Conclusions}

The existence of massive, $\sim 30\,M_\odot$, BHs formed from stars born at solar metallicity has multiple observational implications for binary systems in the context of GW detections, XRBs and astrometric binaries. 
These massive BHs have been observed since the first detection of GWs, GW150914,\cite{2016PhRvL.116f1102A} which were previously interpreted to have originated from the evolution of sub-solar metallicity stars.\cite{2016ApJ...818L..22A}
Such conclusions were based on rapid BPS models of BBH formation which, in contrast to the models presented here, do not predict such massive merging BBHs at solar metallicity as these models overpredict the expansions and consequently the mass loss of massive stars ($\gtrsim 50\,M_\odot$). 
One would also expect that the predicted existence of such massive BHs at solar metallicity would imply their potential observability in XRBs.\cite{2022arXiv220905505M} If confirmed, a potential candidate of such massive BH is the one harboured in the galactic XRB MAXI J1631-479 with its updated mass estimate in the range of $15\,M_\odot$ to $45\,M_\odot$.\cite{2022arXiv221205293R}
Alternatively, future astrometric observations of detached wide binaries in the Milky Way\cite{2019Sci...366..637T,2021MNRAS.504.2577J,2023MNRAS.518.1057E} and globular clusters\cite{2019A&A...632A...3G,2022MNRAS.511.2914S} will be another potential tool to indirectly observe these massive BHs.\cite{2017ApJ...850L..13B,2021arXiv211005979C} The discovery of these massive BHs in binary systems at solar metallicity will be a new probe into the evolution of massive stars. Such discovery, built upon model predictions made here, has the potential to broadly impact astrophysics, from our theoretical understanding of galaxy evolution to the formation of heavy elements.

This study investigated BH formation from single and binary evolution at solar metallicity. 
In contrast to commonly used stellar models in rapid BPS, we showed that stellar binary models that self-consistently account for the feedback of stellar wind mass loss and binary interactions onto stellar evolution lead to the formation of more massive merging BHs than previously thought.
Our fiducial \texttt{POSYDON} single stellar model predicts that massive stars ($M_\mathrm{ZAMS} \gtrsim 50\,M_\odot$) do not expand to red supergiant radii but rather remain more compact. 
Additionally, stars with ZAMS masses above $80\,M_\odot$ reach the WR phase while still burning hydrogen in their cores. 
These massive stars will experience a longer WR phase compared to lower mass stars, forming lower mass BHs and setting the maximum BH mass at solar metallicity at $\simeq 35 \, M_\odot$ ($\simeq 31.5 \, M_\odot$ when accounting for LBV-type winds).
Even though our model of single stellar evolution predicts the existence of massive BHs at solar metallicity, our BPS model predicts a challenging to discover intrinsic fraction ($\sim 1\%$) of merging BBHs with masses above $30\,M_\odot$. 
When accounting for the LIGO's detectors selection effects at design sensitivity the fraction of massive BHs above $30\,M_\odot$ at solar metallicity increases to $\sim 3\%$ for around one detection per year from binaries in the metallicity range $[0.5Z_\odot,2Z_\odot]$. 
%The relative small fraction of these massive BHs compared to lower mass BHs at solar metallicity originates from reduced expansion of their progenitor stars which cause the massive binaries to avoid a mass transfer phase before the formation of the first-born BH for binaries with initial periods of a few days. 
%Instead, lower mass binaries with initial periods of up to XXX would experience a mass transfer phase leads to their orbits to efficiently shrink leading to the formation of a close binary which evolves into a merging BBH system. 
Finally, we find that binaries with primary stars of initial mass $M_\mathrm{ZAMS}\gtrsim 50\,M_\odot$ and initial period of a few days may undergo episodes of Case-A Roche-lobe overflow mass-transfer, a stable contact phase or avoid mass transfer at all, keeping their orbit close and result to mildly spinning first-born BHs ($\chi_\mathrm{BH1} \lesssim 0.2$) when efficient angular momentum transport is assumed. This predicted BH spin signature is in agreement with the analysis of the observed population of merging BBHs.\cite{2022ApJ...937L..13C} 

Our results demonstrate the importance of the self-consistent modeling of mass loss in BPS models, which have been shown here to be crucial for the interpretation of the origin GW sources, XRBs and astrometric BHs.

\section*{Methods}

\subsection*{Single stellar models}

In this study, we generate a grid of \texttt{MESA}\cite{2011ApJS..192....3P,2013ApJS..208....4P,2015ApJS..220...15P,2018ApJS..234...34P,2019ApJS..243...10P} single star simulations with ZAMS masses in the range $[8,150]\,M_\odot$ with resolution $\Delta M_\mathrm{ZAMS} = 0.25\,M_\odot$. 
Our default model follows the stellar model assumptions of \texttt{POSYDON}.\cite{2022arXiv220205892F} The assumed wind mass loss scheme is known as the \emph{Dutch wind scheme} and includes: the Vink+01\cite{2001A&A...369..574V} line-driven winds for stars with effective temperatures above 11,000 K, a linear transition phase to the de Jager+88\cite{1988A&AS...72..259D} dust-driven winds for cooler stars with effective temperatures below 10,000~K, and Nugis\&Lamers00\cite{2000A&A...360..227N} winds for WR stars defined as having effective temperatures above 11,000~K when the surface ${}^1\mathrm{H}$ abundance is below the value $0.4$. 

The default \texttt{POSYDON} stellar model assumes there is no LBV-like wind enhanced mass loss when stars evolve through the Humphreys--Davidson limit.\cite{1994PASP..106.1025H} 
This choice was made given the uncertain estimates of LBV-type wind mass losses.\cite{2017RSPTA.37560268S}
To verify how our results depend on the assumed lack of LBV-type wind, we consider a model variation including LBV-like winds following Belczynski+10.\cite{2010ApJ...714.1217B} In practice, 
the wind mass loss during the LBV-wind phase (shown in the top-right panel of Figure~\ref{fig:winds}) is enhanced to $10^{-4}\, M_\odot \mathrm{yr}^{-1}$ for stars with both
$ L > 6\times 10^{5}\,L_\odot$ and $R/R_\odot (L /L_\odot )^{1/2} > 10^5$  compared to default model shown in the left plot of Figure~\ref{fig:default_model}. Additionally, to illustrate the significance of winds to the evolution of massive stars ($M_\mathrm{ZAMS}\geq 50\,M_\odot$), we consider an additional model variation where no stellar-wind mass loss are implemented. This alternative model, presented in the Supplementary information, shows that neglecting the feedback of mass loss into stellar evolution leads to the expansion of stars to supergiant radii independently from their initial mass.

The stellar-evolution models are stopped when core carbon is depleted, after which we assume their collapse is imminent. 
We evaluate the core-collapse of the massive stars into BHs following the Patton\&Sukhbold22\cite{2020MNRAS.499.2803P} N20 engine prescription, as implemented in \texttt{POSYDON}.\cite{2022arXiv220205892F} 
To verify how our results depend on the assumed core-collapse prescription, we also consider the commonly used Fryer+12\cite{2012ApJ...749...91F} Delayed prescription, from now on referred to as (Fryer+12-delayed), as implemented in \texttt{POSYDON}.\cite{2022arXiv220205892F} 

\subsection*{Binary population synthesis model}
We use the \texttt{POSYDON} framework, to generate a BBH population-synthesis model based on grids of detailed \texttt{MESA} stellar structure and binary evolution simulations (revision 11701). 
We simulated a total of 50 million massive binary systems following the default assumptions made in \texttt{POSYDON} v1.0 instrument paper.\cite{2022arXiv220205892F} 
This set of default assumptions is not the result of specific model calibration. Rather, model assumptions were made, given our best current theoretical and observational understanding of stellar and binary physics processes. 
\texttt{POSYDON} generates a synthetic catalog of merging BBHs at solar metallicity. 
To obtain the underlying (intrinsic) population of merging BBHs at solar metallicity in the Universe, we convolve the synthetic population with the \texttt{IllustrisTNG}\cite{2015A&C....13...12N} star formation rate (SFR) as implemented in Bavera+.\cite{2020A&A...635A..97B,2021A&A...647A.153B,2022A&A...657L...8B} 
As our population-synthesis model is evaluated at solar metallicity, we only account for the SFR in the metallicity range $[0.5Z\odot,2Z\odot]$.
To obtain the detectable BBH population observed by Advanced LIGO\cite{2015CQGra..32g4001L} at design sensitivity,\cite{2018LRR....21....3A} we account for the selection effects of the GW detector following the methodology and implementation of Barrett+18.\cite{2018MNRAS.477.4685B} In practice, we compute the detection probability of a merging BBH system given the BH mass components and the redshift of merger. Here, we ignored the impact of BH spin on detectability as our intrinsic merging BBH population is dominated by slowly spinning BHs and the impact of BH spins on detectability is minor compared to the BH mass selection effects.\cite{2018PhRvD..98h3007N}
The optimal signal-to-noise ratio for a face-on source is computed for a single detector using the sensitivity above with GW waveforms from \texttt{lalsuite}.\cite{lalsuite} The optimal signal-to-noise ratio is then convolved with the antenna pattern function distribution,\cite{1993PhRvD..47.2198F} which allows us to estimate the probability of detection.
In our model we assume that signals are detected if their single-detector signal-to-noise ratio threshold exceeds a threshold value of 8.\cite{2016PhRvD..93d2006A}

\section*{Acknowledgements}
This work was supported by the Swiss National Science Foundation Professorship grant (project numbers PP00P2\_176868 and PP00P2\_211006). 
The \texttt{POSYDON} project is supported primarily by two sources: a Swiss National Science Foundation Professorship grant (PI Fragos, project number PP00P2\_176868) and the Gordon and Betty Moore Foundation (PI Kalogera, grant award GBMF8477). 
%The collaboration was also supported by the European Union's Horizon 2020 research and innovation program under the Marie Sklodowska-Curie RISE action, grant agreements No 691164 (ASTROSTAT) and No 873089 (ASTROSTAT-II).
CPLB acknowledges support from the University of Glasgow.
KK acknowledges support from the Federal Commission for Scholarships for Foreign Students for the Swiss Government Excellence Scholarship (ESKAS No. 2021.0277), and the Spanish State Research Agency, through the María de Maeztu Program for Centers and Units of Excellence in R\&D, No. CEX2020-001058-M.
K.A.R.\ also thanks the LSSTC Data Science Fellowship Program, which is funded by LSSTC, NSF Cybertraining Grant No.\ 1829740, the Brinson Foundation, and the Moore Foundation; their participation in the program has benefited this work.
All figures were made with the open-source Python module \texttt{Matplotlib}.\cite{Hunter:2007} 
This research used the Python modules \texttt{Astropy},\cite{price2018astropy} \texttt{iPython},\cite{PER-GRA:2007} \texttt{Numpy},\cite{harris2020array}
\texttt{Pandas},\cite{reback2020pandas} and \texttt{SciPy}.\cite{2020SciPy-NMeth}

\section*{Authors contributions}
All authors contributed to the work presented in this paper. SSB lead the writing with substantial contributions of TF, EZ, JJA, VK, CPLB, and MK. The results presented in this study were obtained using the open-source software \texttt{POSYDON}, which was developed by SSB, TF, JJA, EZ, AD, KK, DM, KAR, PMS, MS and ZX.

\section*{Competing interests}
The authors declare no competing interests.

\bibliography{main}

\newpage
\appendix

\section*{Supplementary information}

\section{No stellar wind model variation}\label{app:LBV}

\begin{figure}[ht]
\centering
\includegraphics[]{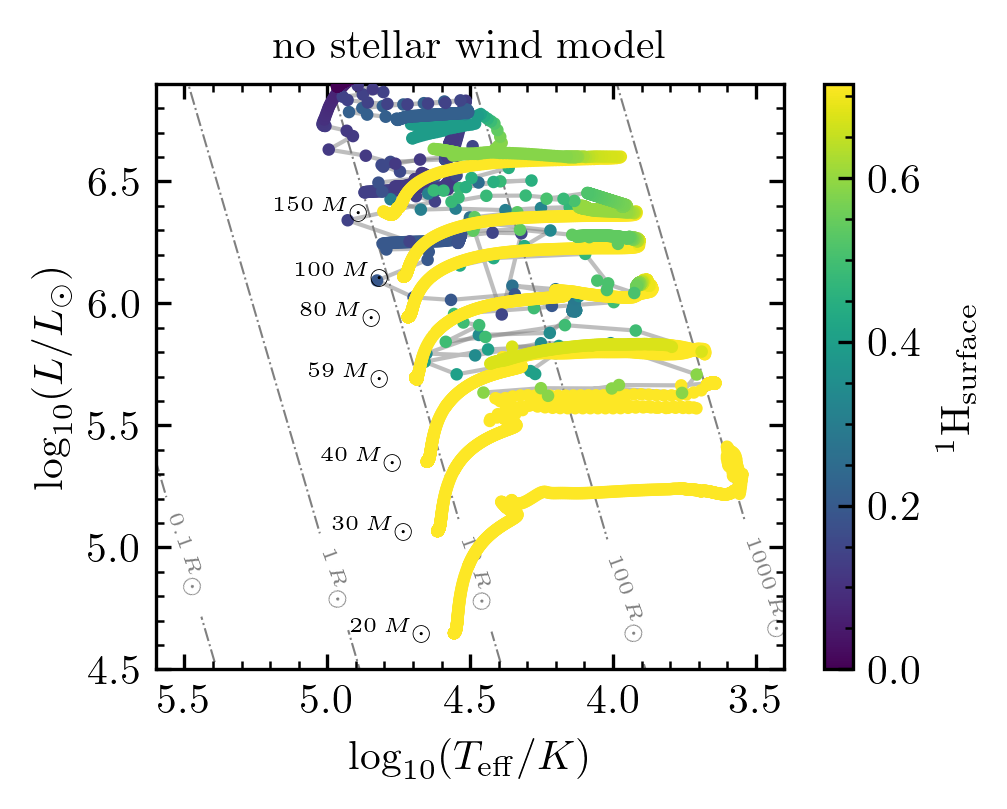}
\includegraphics[]{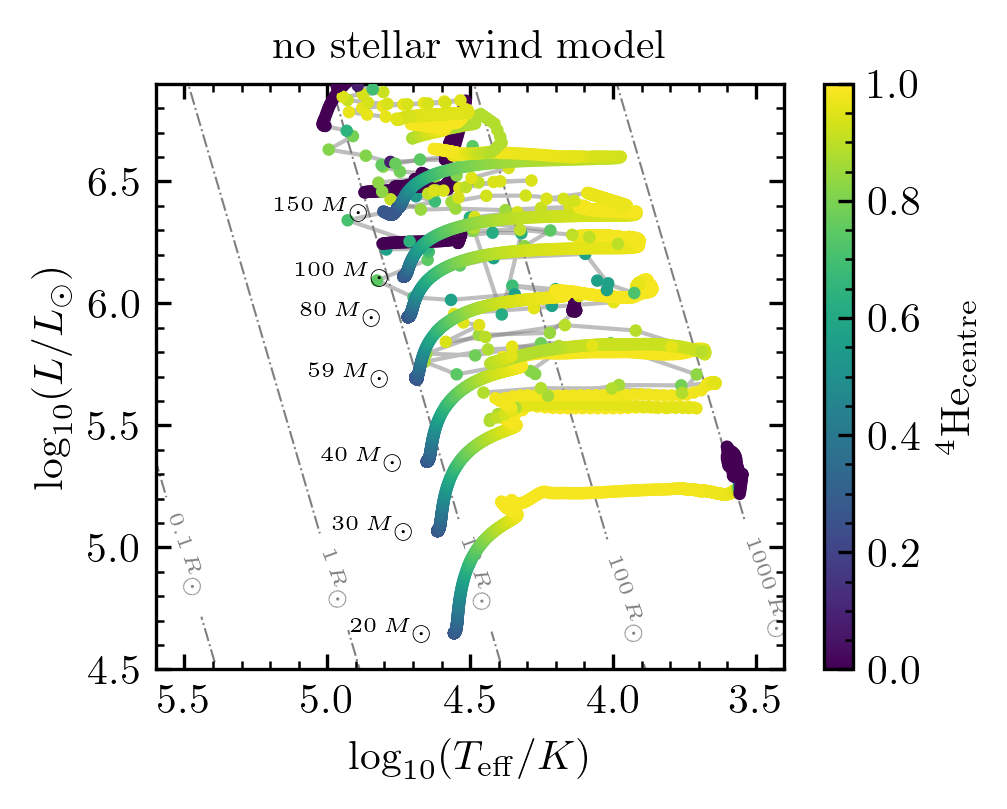}
\caption{Hertzsprung--Russell diagram for stars with masses between $20\,M_\odot$ and $150\,M_\odot$. 
We show a variation of our stellar models without any stellar wind mass loss. 
The instantaneous surface hydrogen abundance (\textit{left}) and centre helium abundance (\textit{right}) are colour-coded in the tracks. In contrast to the default model, we find that not accounting for the stellar wind feedback onto stellar evolution in the stellar models cause massive stars ($\gtrsim 50\,M_\odot$) to expand up to $\sim 1000\,R_\odot$.}
\label{fig:no_wind_model}
\end{figure}

We consider an illustrative model variation where there is no mass loss due to stellar winds. 
Figure~\ref{fig:no_wind_model} shows the stellar evolutionary tracks of such model showing the surface ${}^1\mathrm{H}$ and centre ${}^4\mathrm{He}$ abundances of the stars during their evolution. 
Similar to the constant-mass stellar tracks implemented in \texttt{SSE},\cite{2000MNRAS.315..543H} we find that independently of their ZAMS mass, these massive stars expand to become red-supergiant stars with radii of $\sim 1000\,R_\odot$. 
This exercise shows the importance of self-consistently modeling stellar-wind mass loss in stellar evolution. 
Failing to do so leads to miscalculation of the structure resulting in bigger cores, altered surface chemistry, and nonphysical binary interactions with important implications for BBH formation.

\section{The origin of spin in first-born black holes}\label{app:grid}

\begin{figure}[ht]
\centering
\includegraphics[]{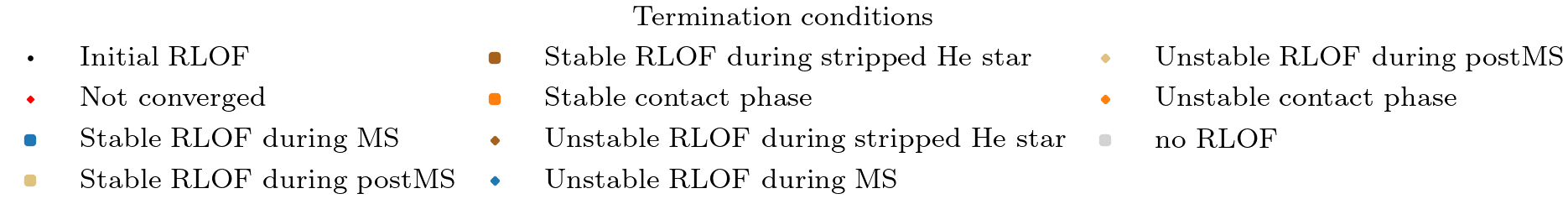}
\includegraphics[]{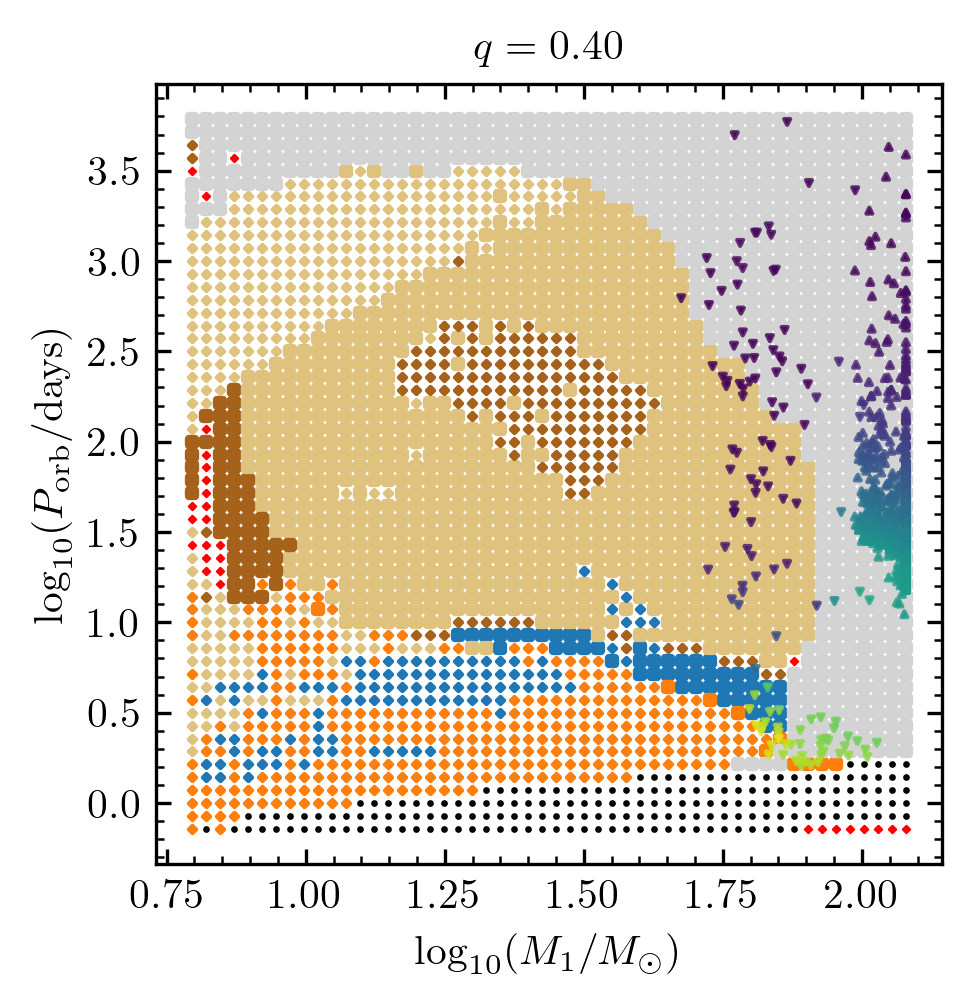}
\includegraphics[]{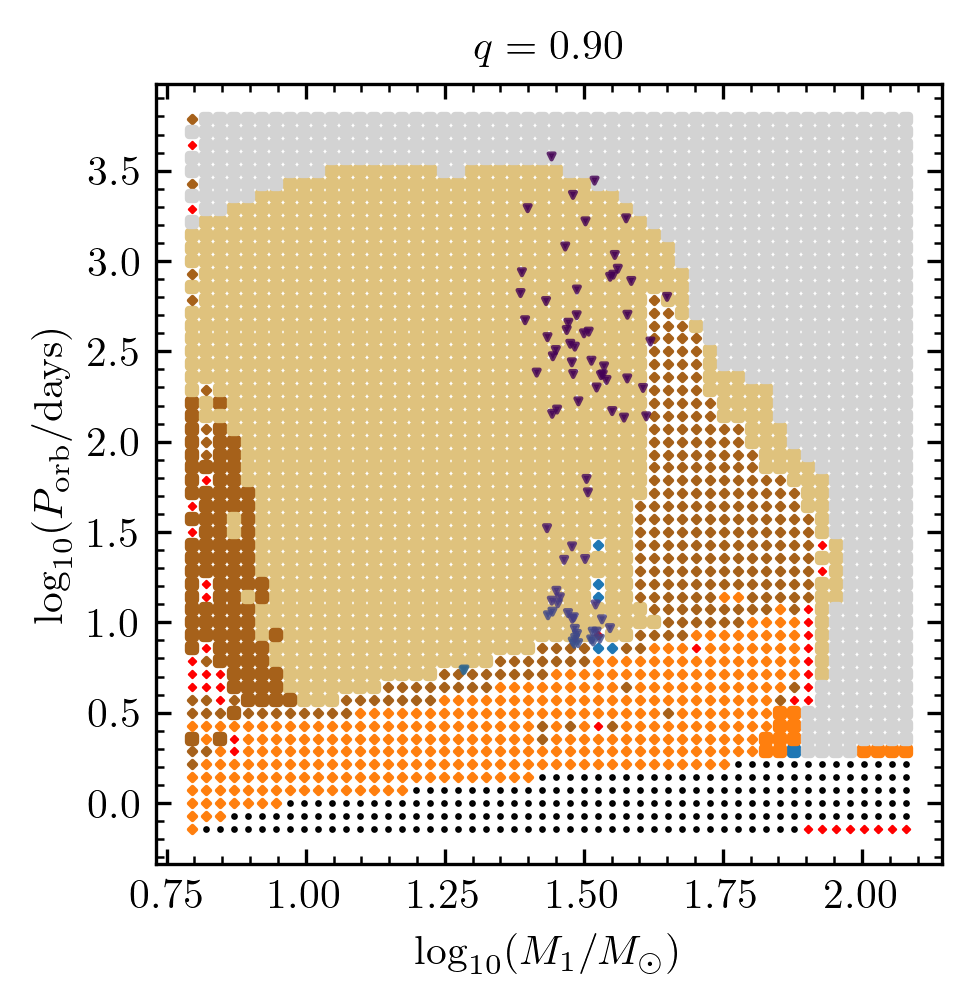}
\qquad
\includegraphics[]{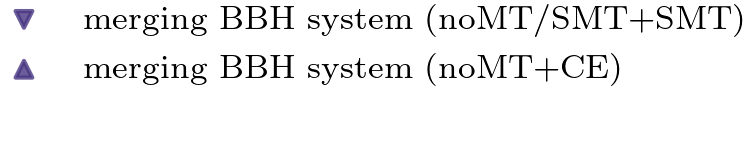}
\quad
\includegraphics[]{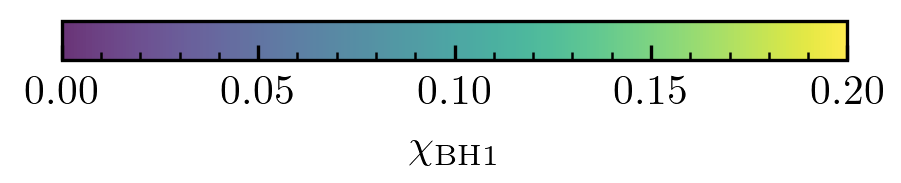}
\caption{
View of two \texttt{MESA} grid slices, for two different values of initial binary mass ratio ($q=0.4$ on the left, $q=0.9$ on the right), from our \texttt{MESA} grid of binary-star models consisting of two stars, initially at ZAMS. 
The different symbols summarize the evolution of each of the models. 
We distinguish between models that experienced stable and no mass transfer (squares), which reach the end of the life of the primary stars, and the ones that stopped during mass transfer due to dynamical instability (diamonds). 
Different colors distinguish the evolutionary phase of the donor star during the latest episode of mass transfer. We also indicate systems that were in initial Roche-lobe overflow (RLOF) at birth and those that stopped prematurely for numerical reasons.  Finally, we overlay with triangle markers binary systems of the \texttt{POSYDON} population synthesis model that further evolve to merging BBHs distinguishing systems that later evolve through an additional stable mass transfer (SMT) or common envelope (CE) according to the legend. The colors of these markers indicate the spin magnitude of the first-born BH according to the colorbar. We see most of mildly spinning BHs to originate in the part of the parameter space of massive primary stars and low initial orbital separations in the mass ratio slice $q=0.4$.
}
\label{fig:grids}
\end{figure}

In Figure~\ref{fig:grids}, we provide two two-dimensional slices of the \texttt{MESA} binary grid of \texttt{POSYDON} consisting of two stars initially at ZAMS. 
We show our simulation outcomes as a function of initial primary star mass $M_1$ and orbital period $P_{\rm orb}$ for a fixed mass ratio $q\equiv M_2/M_1$, where $M_2$ is the secondary initial star mass.
In the left panel we show one example of a mass ratio $q=0.4$, and on the right a more-equal mass ratio $q=0.9$ slice. 
Each point in the panels represents a separate \texttt{MESA} binary simulation from our grid where different markers shapes and colors distinguish different binary evolution phases prior to the formation of the first-born compact object. 
An exception to this are triangular markers which are overimposed onto the grid and represent binary systems from our fiducial \texttt{POSYDON} population synthesis study, which will evolve to become a merging BBH.
These systems are the subsample of merging BBHs with ZAMS mass ratios corresponding to the mass ratios of $0.4\pm 0.025$ and $0.9\pm 0.025$, respectively. 
The colour of these triangular markers indicate the spin of the first-born BH.
We distinguish binaries that will later evolve through a stable mass-transfer phase from the ones that later evolve through a common-envelope phase; the latter are only present in the mass ratio slice of $q=0.4$.
At the bottom of the left panel, at short orbital periods and for massive stars $M_\mathrm{1} \gtrsim 60 \, M_\odot$, we see binaries that either undergo a stable mass transfer, a contact stable phase or, alternatively avoided any mass transfer phase and formed mildly spinning first-born BHs ($\chi_\mathrm{BH1} \leq 0.2$). 
Figure~\ref{fig:grids} shows that this evolution would not be possible at similar initial orbital periods and smaller primary masses as the increased maximal expansion of these stars would lead to unstable mass transfer, followed by a stellar merger. Finally, this evolutionary pathway is not present at more-equal mass ratios because the more massive stellar companion would not expand to become a supergiant star to initiate a mass transfer episode after the formation of the first-born BH.

\end{document}